\documentclass[11pt]{iopart}
\usepackage{iopams}
\usepackage{hyperref}
\usepackage{setstack}
\usepackage{graphics}
\usepackage{epsf}
\usepackage{graphicx}

\begin{document}

\title{Magnetic trapping of a cold Rb-Cs atomic mixture}

\author{M L Harris, P Tierney and S L Cornish}

\address{Department of Physics, Durham University, South Road, Durham DH1 3LE, UK}

\begin{abstract}
We present an apparatus for the study of an ultracold gaseous atomic mixture of $^{133}$Cs and $^{87}$Rb. The mixture is prepared using a double
magneto-optical trap (MOT) system in which a two-species pyramid MOT acts as a source of cold atoms for a `science' MOT. Measurements of the
interspecies trap loss rate coefficients $\beta_{\rm{RbCs}}$ and $\beta_{\rm{CsRb}}$ in the science MOT are reported. After the initial MOT
phase, atoms in the mixture are optically pumped into the magnetically trappable $|F\!=\!3,m_F\!=\!-3\rangle$ and $|F\!=\!1, m_F\!=\!-1\rangle$
states of Cs and Rb (respectively) and loaded into an Ioffe-Pritchard magnetic trap. We demonstrate a novel technique for limiting the
interspecies loss rate in the science MOT by spatially separating the two trapped atom clouds, which greatly enhances the number of atoms which
can be loaded into the magnetic trap.
\end{abstract}

\ead{s.l.cornish@durham.ac.uk}


\maketitle

\section{Introduction}
Quantum degenerate mixtures of two or more atomic species exhibit a wide range of fascinating phenomena not present in single-species systems.
The balance between interspecies and intraspecies atomic interactions can lead to phase separation of the mixture \cite{Pu} and can trigger a
collapse even when the two species are separately stable \cite{Modugno02}. When confined in an optical lattice potential, binary atomic mixtures
display a rich spectrum of quantum phases \cite{Lewenstein}. Such mixtures offer a means of forming ultracold heteronuclear molecules via
photoassociation \cite{Jones} or Feshbach resonances \cite{Kohler}. This in turn facilitates the creation of dipolar quantum gases
\cite{Santos00} and may provide an alternative approach to quantum information processing \cite{DeMille}.  On the technical side, species with
unfavourable elastic to inelastic collision ratios, \textit{e.g.}$^{41}$K \cite{Modugno}, may be cooled via elastic collisions with a second
evaporatively cooled species. This method of sympathetic cooling has proved essential for cooling fermionic quantum gases to degeneracy
\cite{DeMarco}, avoiding the restriction imposed by the suppression of s-wave collisions in single-species fermionic gases due to the Pauli
exclusion principle. Bosons have been used to sympathetically cool fermions to degeneracy in several experiments \cite{Truscott}-\cite{Roati},
permitting the study of mixtures in which the component species obey different quantum statistics. More recently binary bosonic mixtures of
$^{85}$Rb-$^{87}$Rb \cite{Bloch, Papp}, $^{87}$Rb-Cs \cite{Anderlini, Haas}, and $^{87}$Rb-$^{41}$K \cite{Catani} have attracted experimental
attention.

The combination of $^{87}$Rb and $^{133}$Cs offers particularly interesting prospects for experiments on quantum degenerate mixtures.  $^{87}$Rb
(hereafter denoted Rb) is the most commonly condensed alkali species due to its favourable ratio of elastic to inelastic collisions.  In
contrast, strong inelastic losses in Cs make it difficult to condense in a magnetic trap \cite{Odelin, Ma}, and use of the
$|F\!=\!3,m_F\!=\!+3\rangle$ state in combination with several stages of optical trapping were required to achieve quantum degeneracy
\cite{Weber}.  The Rb scattering length is essentially independent of magnetic field, with only narrow Feshbach resonances being present at
accessible magnetic fields \cite{Stefan}.  For Cs, a rich Feshbach structure containing both narrow and broad resonances has already been
exploited to produce cold molecular samples of dimers \cite{Herbig} and Efimov trimers \cite{Kraemer} and to precisely tune the scattering
length around a broad zero-crossing \cite{Gustavsson}.

Unlike some mixtures of alkali metals (particularly those containing Li), the magnetic moment-to-mass ratios for Cs and Rb atoms in the
$|F\!=\!3,m_F\!=\!\pm 3\rangle$ and $|F\!=\!1, m_F\!=\!\pm 1\rangle$ states (respectively) are very similar, differing by less than 2$\%$ for
magnetic fields below $\sim86$\,Gauss.  This ratio determines the acceleration of an atom due to a spatially varying applied magnetic field.
Hence, the very similar ratios for Rb and Cs mean that their oscillation frequencies in a harmonic magnetic trap differ by just 1$\%$ over this
field range.  The displacements of the Rb and Cs clouds due to gravity (`gravitational sag') are thus nearly identical.  This leads to good
spatial overlap of the clouds even at sub-microkelvin temperatures in a weak magnetic trap, making Rb an attractive species with which to
sympathetically cool Cs.  Intriguingly, sympathetic cooling of Cs with Rb may offer a route to quantum degeneracy in magnetically-trapped Cs,
particularly if the magnetic trap is combined with an optical `dimple' potential \cite{Stamper-Kurn}. Using a magnetic trap as the reservoir for
loading this dimple rather than a large volume optical dipole trap \cite{Weber} would greatly simplify the necessary experimental apparatus and
would also give access to the equally rich Feshbach structure in the $|F\!=\!3,m_F\!=\!-3\rangle$ state of Cs.

In this paper, we describe an apparatus for the study of an ultracold gaseous atomic mixture of Rb and Cs. A two-species pyramidal
magneto-optical trap (pyramid MOT)\cite{Lee, Arlt} acts as source for loading slow Rb and Cs atoms into a second `science' MOT.  Atoms collected
in the science MOT are subsequently optically pumped into the $|F\!=\!3,m_F\!=\!-3\rangle$ and $|F\!=\!1, m_F\!=\!-1\rangle$ states and
transferred into a magnetic trap.  Each stage of this process was independently optimised for each species, and results are presented for both
one-species and two-species operation of the two MOTs and the magnetic trap.  A novel technique for suppressing the strong light-assisted
interspecies inelastic collisions in the MOT by spatially separating the two trapped atomic clouds was developed.  This technique allows us to
control the Rb and Cs atom numbers in the science MOT over a wide range, making the modified MOT an excellent starting point for loading the
mixture into the magnetic trap for further study.

\section{Experimental apparatus}

\begin{figure}[tbh]
\begin{center}
\includegraphics*{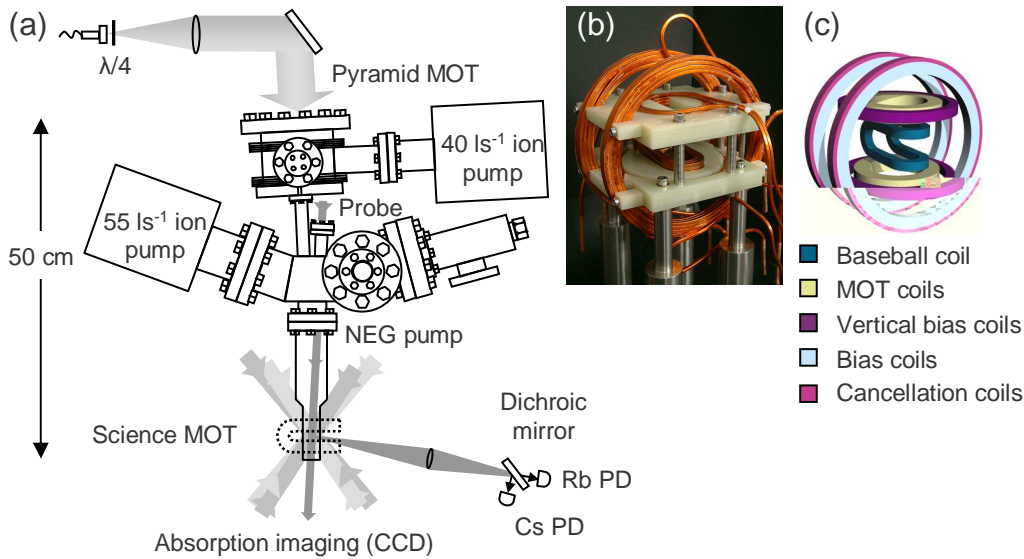}
\end{center}
\caption{\label{layout}(a) (Colour online) Schematic diagram of the two-species double-MOT apparatus. A differential pumping tube aligned at $\sim3^{\circ}$ relative to the science cell connects the two chambers. A pyramid MOT generated by a single beam of 780~nm and 852~nm light acts as a two-species cold atom source for the two overlapped six-beam science MOTs.  The intensity and alignment of the four MOT beams in the plane of the table are controlled independently for each species. Each beam contains both cooling and repumping light. In the vertical direction, the cooling and repumping light for both Rb and Cs are aligned along a single beam path (not shown). Fluorescence from the two-species MOT is collected on separate photodiodes for Rb and Cs using a dichroic mirror. Absorption imaging onto a CCD camera provides spatial information about the trapped clouds. Dotted lines indicate the position of the magnetic trap. (b) Photograph and (c) schematic diagram of the magnetic trap assembly. The coils in (c) are labeled in order of increasing distance from the trap centre.}
\end{figure}

The experiment employs a two-species double MOT arrangement to collect and cool the atomic mixture in an ultra-high vacuum (UHV) environment.
The required laser light is derived using the optical setup presented schematically in Figure~\ref{opticschematic}. Light from this setup is
transferred via optical fibres to a second optical table housing the vacuum apparatus, shown in Figure~\ref{layout}. The key components of the
apparatus are described in detail below.

\subsection{Vacuum apparatus}
The vacuum apparatus for the two-species double MOT consists of a two-chamber system connected by a differential pumping tube. A pyramid MOT
serves to initially collect and cool the atoms from a background vapour. A gap at the apex of the pyramid leads to a flux of cold atoms into the
UHV chamber, where they are captured in the science MOT. The pyramid MOT chamber is pumped by a single 40\,ls$^{-1}$ ion pump (Varian) and
houses commercial dispensers (SAES) for Rb and Cs. In total six dispensers, three for each species, are spot-welded to two separate electrical
feedthroughs. The vapour pressure of each species is controlled independently by running separate currents through one of the Rb and one of the
Cs dispensers. The differential pumping tube comprises two sections: a 1.7\,cm tube with an inner diameter of 5\,mm, followed by a 8.7\,cm tube
with an inner diameter of 16\,mm. The combined conductance is approximately 0.7\,ls$^{-1}$. The tube is aligned at a $\sim3^{\circ}$ angle
relative to the axis of the science cell in order to create space for a 34\,mm outer diameter viewport at a similar angle in the opposite
direction (see Figure~\ref{layout}). This provides optical access for absorption imaging along the axis of the cell.

The rectangular science cell is made from 2\,mm thick fused silica and has internal dimensions of $20\times20\times83\,\mbox{mm}$. A cylindrical
graded index section increases the overall length of the cell from the flange to $\sim21\,\mbox{cm}$. The cell is pumped by a 55\,ls$^{-1}$ ion
pump (Varian) and a non-evaporable getter (NEG) pump (SAES).

\subsection{Laser setup}
\begin{figure}[tbh]
\centering
\includegraphics*{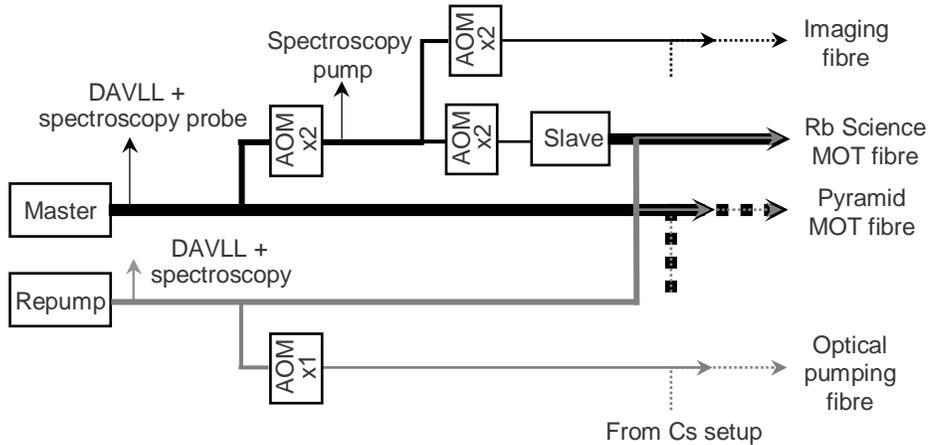}
\caption{\label{opticschematic} Schematic diagram of the optical layout for Rb. Three lasers provide light for cooling, repumping, imaging and
optical pumping in the double-MOT system. The Cs layout is similar.  Line thickness indicates relative beam power. Acousto-optic modulators
(AOMs) in double pass (x2) and single pass (x1) configurations allow independent control of all frequency detunings.  Larger shifts in frequency
required for optical pumping and depumping stages are obtained by additionally applying a dc voltage offset to the dichroic atomic vapour laser
lock (DAVLL) signal in order to offset the laser lock point.  All light is transferred from the laser table to the experiment via optical
fibres.  For the pyramid MOT, optical pumping, and imaging light, dichroic optics are used to combine 780~nm (solid lines) and 852~nm (dashed
lines) beams before transmitting both wavelengths down single fibres.}
\end{figure}

A system of six diode lasers (three for each species) provides the light required for the experiment. Figure \ref{opticschematic} contains a
schematic diagram of the system for Rb; the Cs layout is similar.  Two commercial `master' lasers (Sacher Lasertechnik Lynx TEC-120) providing
$\sim$150~mW at 780~nm and 852~nm generate light for cooling in the pyramid MOT, imaging, and injecting two homebuilt `slave' lasers which
provide cooling light for the science MOT.  Other homebuilt extended cavity diode lasers supply repumping light for both MOTs, and light for
optical pumping.

Both the master lasers and the repump lasers are locked using the dichroic atomic vapour laser lock (DAVLL) method \cite{Corwin,Fred}. Saturated
absorption/hyperfine pumping spectroscopy \cite{MacAdam,Dave} acts as a frequency reference for the lock.  The frequency of the Rb master laser
is stabilised at the optimum detuning for the pyramid MOT. To do this, the frequency of the pump beam in the reference spectra is shifted by
$-231.3(1)$~MHz relative to the probe beam using a double-passed AOM. This allows the large $F=2 \rightarrow F^{'}=2,3$ crossover peak at a
detuning of approximately $-133.5\,\mbox{MHz}$ to be used as a frequency reference. For the Rb repump laser the hyperfine peak associated with
the repumping transition is used as a reference.  To improve the stability of the lock, the magnetic field required to produce the DAVLL signal
is generated by a solenoid and the ohmic heating is used to increase the vapour pressure in the cell \cite{Danny}. The Cs lasers are locked
following a similar scheme.

Additional double-passed AOMs are used to diffract some light from the spectroscopy AOM back to the detunings required for the science MOT and
absorption imaging (Figure \ref{opticschematic}).  A single-passed AOM is used to rapidly switch off the optical pumping light.  Larger
frequency shifts required for optical pumping (depumping) are obtained by additionally applying a dc voltage offset to the DAVLL signal used for
locking the repump (master) laser, thereby offsetting the laser lock point.

Light from the optical table is transmitted to the experiment via a total of five polarization maintaining fibres. For the pyramid MOT, cooling
and repumping light are combined into a single beam using a polarizing beamsplitter. The 780~nm and 852~nm beams are then overlapped using a
dichroic mirror which transmits (reflects) light at 780~nm (852~nm).  The resulting four-frequency beam is coupled into a single fibre, with the
polarizations of the cooling and repumping beams aligned to the two orthogonal axes of the fibre.  Dichroic mirrors are also used to couple both
wavelengths into the fibres which transmit light for imaging and optical pumping. Light for the Rb and Cs science MOTs is coupled into separate
fibres for each wavelength.

\subsection{Pyramid MOT}

The pyramid MOT provides a compact, robust two species atom source. The pyramid is constructed in two steps. Firstly, two rectangular
mirrors are mounted at 90 degrees to each other to form a `V' shaped trough. Into this trough, two prisms are then mounted so as to close off
the two ends of the trough. The four optics then form an open square based pyramid with a small ($\sim2.4\times3.0\,\mbox{mm}$) opening at the apex. The face of each optic that points into the pyramid volume has a high reflection coating. The base of the pyramid is $60\times60\,\mbox{mm}$. The gap at the apex of the
pyramid is aligned with the differential pumping tube leading into the UHV region of the vacuum system.

Light from the pyramid MOT fibre ($\sim50$~mW of cooling and $\sim2$~mW of repumping power at both wavelengths) passes through a single
achromatic $\lambda$/4 plate and is collimated (1/$e^2$ radius = 20.4(4)~mm) before entering the vacuum chamber.   Aligning the circularly
polarized pyramid MOT beam onto the mirror assembly produces three orthogonal pairs of beams with the polarizations of a standard six-beam MOT,
as described in \cite{Arlt}.

The magnetic field gradient required for the pyramid MOT is produced by a pair of water-cooled circular coils operated in the anti-Helmholtz
configuration and wound directly onto the vacuum system.  Additional coils provide a bias field, allowing the position of the magnetic field
zero to be adjusted. When the zero-field point is located directly in front of the pyramid apex, the pyramid MOT acts as a cold atom source for
the science MOT.

\subsection{Science MOT}
Atoms in the cold atomic beam generated by the pyramid MOT are captured in two six-beam science MOTs. In the horizontal plane there are eight
beams (four per species), the intensity, alignment and polarisation of which can all be independently adjusted.  The 780~nm and 852~nm beams in
this plane are separated by $\sim7^{\circ}$. In the vertical direction the 780~nm and 852~nm light is aligned along a common beam path. All of
the beams contain both cooling and repumping light. Having independent adjustment in the horizontal plane means that each single-species MOT can
be independently optimised without affecting the performance of the MOT for the other species.

Light from the Rb and Cs science MOT fibres is collimated to a 1/$e^2$ radius of 10.9(1)~mm.  The beams are subsequently truncated by passing
through 20~mm square polarizing beamsplitter cubes.   A pair of anti-Helmholtz coils mounted on a framework outside the science cell (Figure
\ref{layout}(b) and (c)) provides the magnetic field gradient.

\subsection{Magnetic trap}
The experiment uses an Ioffe-Pritchard type `baseball' trap to generate 3D confinement with a controllable, non-zero bias field.  The baseball
consists of nine windings of Kapton-coated square copper tubing (outer diameter 3.175~mm with a 2.00~mm round hole down the centre).  This coil
alone produces a radial gradient of 1.087(4) Gcm$^{-1}$A$^{-1}$, an axial curvature of 0.154(2) Gcm$^{-2}$A$^{-1}$, and a bias field of
0.3410(3) GA$^{-1}$.  Additional bias and cancellation coils (nine and three turns, respectively, of tubing with an outer diameter of 4~mm and
an inner diameter of 2.75~mm) in the Helmholtz configuration are used to either enhance or cancel the bias field produced by the baseball.
Vertical bias coils will be used in future experiments on optically trapped atoms in combination with a levitation field gradient produced by
the MOT coils. Figures \ref{layout}(b) and \ref{layout}(c) show the arrangement of the coils.

\subsection{Diagnostics}
Both fluorescence detection and absorption imaging are used to probe the atomic mixture. Fluorescence from the MOT is focused using a single
lens and directed onto separate photodiodes for Rb and Cs light using a dichroic mirror. The dichroic reflects 852\,nm light and transmits
780\,nm light. A small amount (1.2(2)$\%$) of 780\,nm light is reflected onto the photodiode monitoring the Cs fluorescence. This residual Rb
fluorescence signal is subtracted from the Cs signal.

Absorption imaging provides spatial information about the atoms in both the MOT and the magnetic trap.  It is also used to calibrate the
fluorescence detection.  The Cs and Rb clouds are imaged onto a frame transfer CCD camera (Andor iXon 885). For the MOT measurements we use the
camera in the standard imaging mode and image only a single species in any one experimental run. However, when imaging the mixture following
release from the magnetic trap, we operate the camera in `fast kinetics' mode. This allows us to record separate images of the Rb and Cs clouds
in a single experimental sequence, whilst still using the entire $8\times8\,\mbox{mm}$ active area of the CCD for each image. In this mode the
charge accumulated in the active CCD during the imaging of the first species is subsequently shifted into a second masked CCD. The vertical
shift speed is 1.916~$\mu$s per row of pixels, leading to a minimum delay of just 1.92~ms when all 1000 rows of pixels are used. The active CCD
is then exposed again and an image of the second species is recorded.  Both images are then read into a computer for processing.  The two-image
fast kinetics sequence is performed three times to obtain the usual set of three images (absorption, probe only, and background) for each
species. The total time required is typically $\sim2$~s, limited by the readout speed of the camera.

\section{Results}

\subsection{Pyramid MOT}
 The gradient and detunings of the single-species pyramid MOTs were optimised in order to maximise the loading rate of the science MOT.  This
 was done by monitoring the fluorescence of atoms captured in the science MOT during the first 10~s of loading.  For Cs, the maximum science MOT
 loading rate of 1.3(3)$\times$10$^7$ atoms~s$^{-1}$ was achieved for a pyramid magnetic field gradient of 8.5(1) Gcm$^{-1}$ and a pyramid MOT
 laser detuning of $\Delta / \Gamma=-3.06(4)$.  The maximum loading rate for Rb was 1.8(3) $\times10^7$ atoms~s$^{-1}$, at the same gradient and
 a detuning of $\Delta /\Gamma=-2.80(3)$ (see Table 1). Here, $\Gamma = 2 \pi \times 5.22$~MHz ($2 \pi \times 6.06$~MHz) is the natural
 linewidth for Cs (Rb), while $\Delta$ is $2 \pi \times$ the laser detuning in MHz .  These optimum settings showed only a weak dependence on the
 second MOT parameters in the vicinity of \textit{their} optimum settings. The atomic flux and therefore the science MOT loading rate are
 strongly dependent on the dispenser current which sets the vapour pressure in the pyramid chamber. Increasing the dispenser current above its
 usual operating value of 3\,A led to a much increased atomic flux, but would potentially degrade the vacuum in the UHV region over time. At
 4\,A, for example, the science MOT loading rate is increased by more than an order of magnitude and is then comparable to the rates
 demonstrated previously for a similar Rb-Cs cold atomic beam source \cite{Lunblad}.

 When the two-species pyramid MOT is operated as a cold atom source, the continuous transfer of trapped atoms into the beam severely limits the density of the atom clouds. Consequently, interspecies cold collisions play little role in the loss of atoms from the pyramid MOT.  This was confirmed by observing that the presence of the second species in the pyramid MOT had no detectable effect on the loading rate of the first species in the science MOT.

 \begin{table}
  \centering
\begin{tabular}{|l|c|c|}
\hline
 \rule{0mm}{4mm} &  $^{87}$Rb & $^{133}$Cs \\ \hline
 \rule[-1mm]{0mm}{5mm} Cooling transition & $5S_{1/2} (F=2) \rightarrow 5P_{3/2} (F^{'}=3)$ & $6S_{1/2} (F=4) \rightarrow 6P_{3/2} (F^{'}=5)$ \\
 \rule[-1mm]{0mm}{5mm} Repumping transition  &  $5S_{1/2} (F=1) \rightarrow 5P_{3/2} (F^{'}=2)$ & $6S_{1/2} (F=3) \rightarrow 6P_{3/2} (F^{'}=4)$ \\
\rule[-1mm]{0mm}{5mm} Pyramid MOT detuning ($\Delta/\Gamma$)  &  -2.80(3) & -3.06(4) \\
\rule[-1mm]{0mm}{5mm} Science MOT detuning ($\Delta/\Gamma$)  &  -1.99(3) & -1.97(4) \\
\rule[-1mm]{0mm}{5mm} $I_{\rm{Pyramid}}$ (mW~cm$^{-2}$) &  49.2(6) & 49.8(8) \\
\rule[-1mm]{0mm}{5mm} $I_{\rm{Science}}$ (mW~cm$^{-2}$) &  11.0(3) & 11.8(3) \\
\rule[-1mm]{0mm}{5mm} MOT gradient (G cm$^{-1})$&  10.1(1) & 10.1(1)  \\
\rule[-1mm]{0mm}{5mm} $N_{\rm{MOT}}$ &  9(1)$\times$10$^8$ & 4(1)$\times10^8$\\
\rule[-1mm]{0mm}{5mm} $<\!\!n_{\rm{MOT}}\!\!>$ (cm$^{-3}$) &  8.5(5)$\times10^8$ & 3.1(3)$\times10^9$\\
\hline
\rule[-1mm]{0mm}{5mm} $N_{\rm{MT}}$ &  6(1)$\times10^8$ & 3(1)$\times10^8$  \\
\rule[-1mm]{0mm}{5mm} Temperature ($\mu$K) &  100(30) & 90(20) \\
\rule[-1mm]{0mm}{5mm} $\nu_{\rm{r}}$ (Hz) &  10.82(1) & 11.29(4) \\
\rule[-1mm]{0mm}{5mm} $\nu_{\rm{ax}}$ (Hz) &  3.878(3) & 4.027(2) \\\hline
\end{tabular}
\caption{Summary of the experimental parameters and performance. The natural linewidth $\Gamma$ is 2$\pi\times$6.06~MHz (2$\pi\times$5.22~MHz)
for Rb (Cs) and $\Delta$ is $2 \pi \times$ the laser detuning in MHz. Numbers, densities and temperatures listed are typical single-species
values. The MOT intensities represent the combined values due to all the beams. Trap frequencies were measured at currents of 150~A for the
baseball coil and 100~A for the bias coils and correspond to the conditions under which the magnetic trap is loaded.}
\label{table:RbCsParameters}
\end{table}

\subsection{Science MOT}
The alignment, detuning, and magnetic field gradient of the Rb and Cs science MOTs were optimised in order to maximize the trapped atom numbers
of each species without the other species present. For Rb, a magnetic field gradient of 10.1(1) G~cm$^{-1}$ and a laser detuning of
$\Delta/\Gamma=-1.99(3)$ produced a MOT containing 9(1)$\times$10$^8$ atoms.  For Cs, 4(1)$\times$10$^8$ atoms were trapped in the MOT at the
same optimised gradient and a detuning of $\Delta/\Gamma=-1.97(4)$.

Unlike in the pyramid MOT, the higher densities in the science MOT mean that inelastic cold collisions can often be the dominant source of loss.
The time evolution of a single-species MOT is modeled according to the rate equation \cite{Weiner03}
\begin{equation}
\frac{dN_{i}}{dt}=L-\gamma N_{i}-\beta_i \int_V n^2_{i}d^3r \label{eq:onespeciesrateequation}
\end{equation}
where $N_i$ is the number of atoms in the MOT, $n_{i}$ is their density, $L$ is the loading rate (\textit{e.g.} from the pyramid MOT), $\gamma$
represents the loss rate due to collisions with the background gas, and $\beta_i$ is the loss rate coefficient for light assisted inelastic
collisions between cold atoms of the same species.  The physical processes involved in such collisions (\textit{e.g.} radiative escape, fine
structure change, and hyperfine structure change) have been extensively reviewed by Weiner \textit{et al.} \cite{Weiner} and will not be
discussed here.

\begin{figure}[tbh]
\begin{center}
\includegraphics*{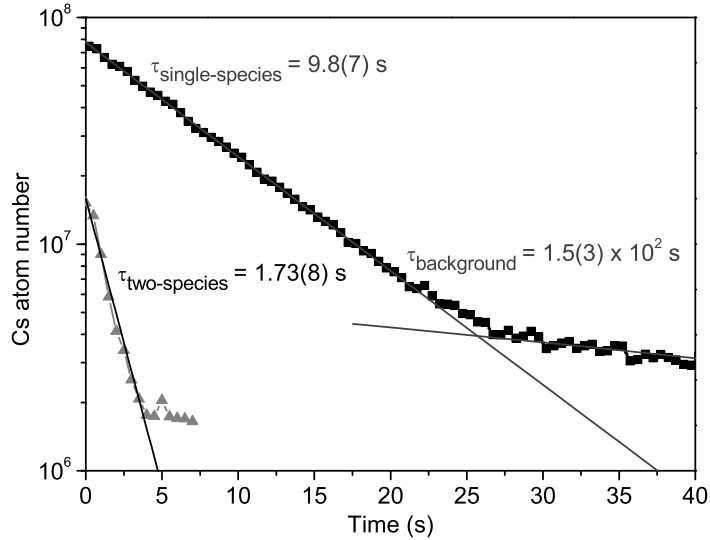}
\end{center}
\caption{\label{decay}Decay of the Cs MOT with (triangles) and without (squares) Rb present. The large value of $\beta_{\rm{CsRb}}$ at the
detunings and intensities used for the MOT measurements leads to a factor of five reduction in the Cs MOT lifetime compared to the
single-species case.}
\end{figure}

During our studies of MOT loss, the magnetic field gradient was changed to 20.8(2)~Gcm$^{-1}$ to ensure that the Cs MOT was located entirely
within a larger Rb MOT.  The detuning of the Cs MOT was also changed to $\Delta/\Gamma\!=-3.49(4)$ to increase the number of trapped Cs atoms at
the higher gradient. To measure the single-species loss rate coefficients, $\beta_{\rm{Rb}}$ and $\beta_{\rm{Cs}}$, the pyramid MOT was switched
off and the science MOT fluorescence observed as the MOT decayed.  Initially, the decay occurs in a density-limited regime, where the MOT loses
atoms and decreases in size without any change in density.  Later, the density begins to decrease, and collisions with the background gas are
the dominant loss mechanism, as illustrated in Figure\,\ref{decay}.

For our experiment, measurements of both species' lifetime in the magnetic trap show that $1/\gamma=$1.5(3)$\times$10$^2$~s.  Data from a series
of absorption images indicated that the MOT was operating in the density-limited regime for the first 200\,s (20\,s) of decay in the Rb (Cs)
MOT.  With the density constant, the solution of Eq.~\ref{eq:onespeciesrateequation} becomes a single exponential decay with a time constant
$\tau=\left(\gamma+\beta_i <\!\!n_i\!\!>\right)^{-1}$, where $<\!\!n_i\!\!>$ is the mean number density. Single exponential fits to the data
over this time period yielded $\beta_{\rm{Rb}}=$ 1.5(2)$\times$10$^{-11}$cm$^3$s$^{-1}$ and $\beta_{\rm{Cs}}$=
2.1(1)$\times$10$^{-11}$cm$^3$s$^{-1}$, where the errors represent statistical variation over four decay measurements per species.  Additional
systematic uncertainties of up to 30$\%$ arise primarily from measurements of atomic cloud sizes.  These values are in good agreement with the
available published data \cite{Gensemer,Sesko}.

For the two-species MOT, Eq. \ref{eq:onespeciesrateequation} becomes \cite{Brazil-Italy}
\begin{equation}
\frac{dN_{i}}{dt}=L-\gamma N_{i}-\beta_i \int_V n^2_{i}d^3r -\beta_{ij} \int_Vn_{i}n_{j}d^3r. \label{eq:twospeciesrateequation}
\end{equation}
where $n_{i,j}$ are density profiles of each species in the science MOT, and $\beta_{ij}$ is the rate constant for the loss of species $i$ due
to collisions with species $j$. Different methods have been used to measure the two-species rate constants in Rb-Cs MOTs
\cite{Brazil-Italy,Lunblad}.  One method uses the decay process described previously.  In this case, Eq. \ref{eq:twospeciesrateequation} becomes
\begin{equation}
\frac{\mathrm{d}N_i}{\mathrm{d}t} = -\left(\gamma+\beta_i <\!\!n_{i}\!\!>+\beta_{ij} <\!\!n_{j}\!\!>F_{ij}\right)N_i
\end{equation}
where the factor $F_{ij}$ represents the (normalised) density of species $j$ weighted by the probability distribution of species $i$.  $F_{ij}$
approaches zero as the two clouds become spatially separated, while two clouds of equal size centred at the same position will have an $F_{ij}$
of unity.  $F_{ij}$ may therefore be thought of as a measure of the relative overlap of the two trapped atom clouds.
Using this method, $\beta_{\rm{RbCs}}$ and $\beta_{\rm{CsRb}}$ were found by comparing the decay rates of Rb and Cs (in the density-limited
regime) in the presence and absence of a second species.  The MOTs were initially loaded to their two-species equilibrium values. The Cs pyramid
MOT beam was then switched off, and the Cs MOT allowed to decay in the presence of cold Rb.  Although the number of Rb atoms increased during
this decay, the position of the Cs MOT near the centre of the much larger Rb cloud meant that the overlap factor $F_{\rm{CsRb}}$ remained
constant, and the Cs decay curve could be described by a single exponential throughout the density-limited regime.  As Figure \ref{decay} shows,
the Cs MOT lifetime is reduced by a factor of five from the single-species value in the presence of Rb.  Averaging over three decay curves
yielded $\beta_{\rm{CsRb}}=1.0(6)\times10^{-10}$cm$^3$s$^{-1}$. Similar measurements of Rb decay in the presence of Cs yielded
$\beta_{\rm{RbCs}}=1.6(4)\times10^{-10}$cm$^3$s$^{-1}$.

The interspecies rate coefficients can also be determined by comparing values at equilibrium for the atom number and the size of trapped clouds
in the single-species MOT to the corresponding equilibrium values with another species present. In this `steady-state' method, Eq.
\ref{eq:twospeciesrateequation} is solved for $\beta_{ij}$ in the presence and absence of a second species, yielding
\begin{equation}
\beta_{ij}= \frac{1}{<\!\!n_{ij}\!\!>N_i}\left[\gamma
\left(\tilde{N}_i-N_i\right)+\beta_i\left(<\!\!\tilde{n}_{i}\!\!>\tilde{N}_i-<\!\!n_{i}\!\!>N_i\right)\right].\label{eq:4}
\end{equation}
where the $\tilde{\:}$ denotes quantities in the single-species MOT (see the Appendix for details of the derivation).

Figure \ref{steadystate} shows a typical loading sequence used to extract the number of atoms present in the single-species and two-species
equilibria.  Initially, no repumping light is present, preventing the MOTs from forming. At $t=0$\,s, the Cs repumping beam is unblocked, and
the Cs MOT loads to its single-species equilibrium level.  At $t=45$~s the Rb repumping beam is unblocked, and collisions with cold Rb atoms
cause rapid losses of $\sim75$\% of the trapped Cs atoms. The loading sequence is then reversed. Absorption images were taken of the Cs and Rb
MOTs at their single-species and two-species equilibrium levels and values for the cloud radii obtained from Gaussian fits to these images.
These radii were used to calculate the average densities appearing in Eq. \ref{eq:4}.
\begin{figure}[tbh]
\begin{center}
\includegraphics*{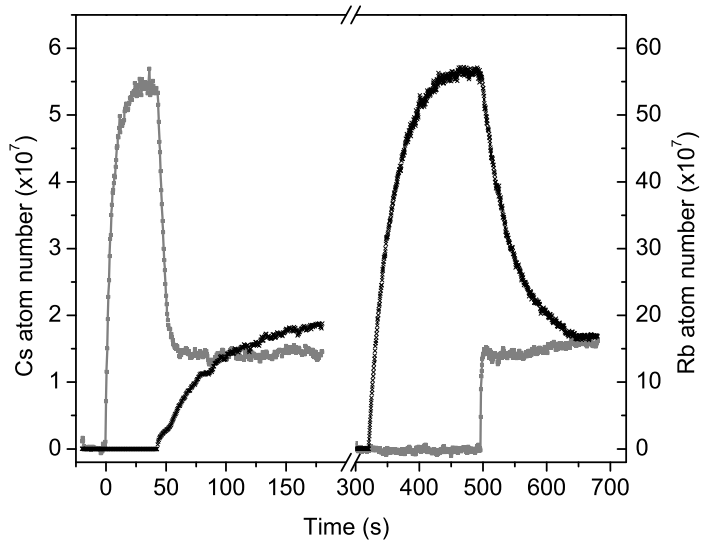}
\end{center}
\caption{\label{steadystate}Loading and loss in the Rb-Cs MOT. Initially, only Cs (grey line) is trapped.  Unblocking the Rb repumping beam
after 45\,s leads to rapid losses of $\sim$~75\% in the Cs cloud. At $t=300$\,s the loading sequence is reversed. Note that the Cs MOT reaches
its two-species equilibrium value within 10\,s after the Cs repumping beam is unblocked, while the Rb cloud decays much more slowly due to the
reduced overlap with the Cs cloud.}
\end{figure}
Based on these steady-state measurements, we found $\beta_{\rm{RbCs}}=2.1(3) \times 10^{-10}$ cm$^3$s$^{-1}$ and $\beta_{\rm{CsRb}}= 9(2) \times
10^{-11}$ cm$^3$s$^{-1}$, where the errors represent statistical variation over 3 (4) data sets for $\beta_{\rm{RbCs}}$ ($\beta_{\rm{CsRb}}$).
Additional systematic uncertainties of up to 30$\%$ arise primarily from measurements of atomic cloud sizes.  These values are in good agreement
with the values from the decay measurements.

\subsection{Displaced science MOT}
The very strong interspecies loss evident from both measurements presents a challenge for optimising the two-species Rb-Cs MOT.  One possible
optimisation strategy would be to alter the alignment and intensity balance of the MOT beams from their single-species optimised values in order
to reduce the overlap of the two clouds. This realignment can be time--consuming, and ultimately one wants the two clouds well overlapped in
order to load both into the centre of the magnetic trap. Additionally, changing the detunings and intensity of the MOT beams can in some cases
be beneficial. Gensemer \textit{et al.} demonstrated that for Rb, shifting the laser detuning by just one natural linewidth can change
$\beta_{\rm{Rb}}$ by up to three orders of magnitude \cite{Gensemer}.  However, as Weiner \textit{et al.} have discussed, the relationship
between detuning, intensity, and trap loss rate is quite complex even for single-species MOTs \cite{Weiner}.  The parameter space is thus very
large, and it is difficult to know \textit{a priori} what effect changes will have on the loss rate.

We have developed an alternative method for reducing the interspecies loss which does not require any deviation from the optimum single-species
MOT parameters.  In this technique, a beam of light (with a 1/e$^2$ radius of $7(1)\,\mbox{mm}$ and a peak intensity of
$0.48(2)\,\mbox{mWcm}^{-2}$) close to resonance with the Rb cooling transition is aligned through the MOT centre during loading.  The effect is
to shift the centre of the Rb MOT by several mm, spatially separating the trapped Rb and Cs clouds. This `push' beam is simply the beam used for
absorption imaging at a higher intensity.  Implementing the displaced MOT thus requires no additional optics or alignment. Optimum conditions
for single-species trapping can be recovered by turning off the push beam, with no need for complex and time-consuming adjustments.

\begin{figure}[tbh]
\begin{center}
\includegraphics*{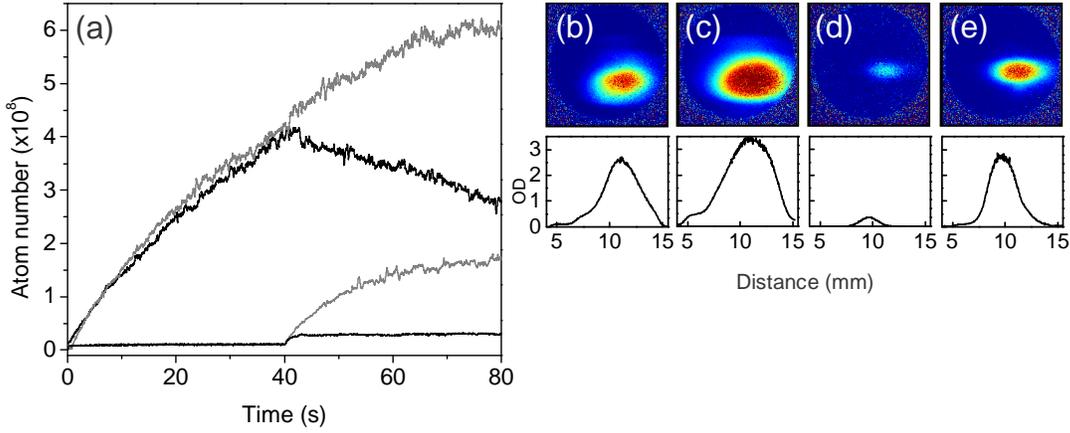}
\end{center}
\caption{\label{pushbeam}Performance of the displaced double MOT. (a) The Cs MOT is switched on after 40~s of loading Rb alone. Without the push
beam (black line), atoms are lost from the Rb MOT due to light assisted interspecies inelastic collisions, and loading of the Cs MOT is greatly
impaired. Turning the push beam on (grey line) displaces the Rb MOT from the Cs MOT loading region, limiting losses due to interspecies
collisions. Absorption images of magnetically trapped Rb clouds loaded from unperturbed (b) and displaced (c) MOTs are also shown. Cross
sections of the clouds in the vertical direction show increases in optical depth and size with the push beam present.  Similar images of Cs
clouds without (d) and with (e) the Rb MOT displaced show even more significant improvement.  All images were obtained with the second species
present in the MOT.  Images are displayed on a common optical depth scale of 0-3 except for the unperturbed Cs cloud, which is on a 0-1 scale.
The cross sections are shown on a common 0-3.5 optical depth scale.}
\end{figure}

Figure \ref{pushbeam}(a) shows the fluorescence of the Rb and Cs MOTs during loading with and without the Rb MOT displaced.  The Rb MOT is
allowed to load to approximately 50$\%$ of its maximum single-species level before the Cs MOT loading is turned on at $t=40$~s.  With the push
beam off, the Cs MOT loading is strongly suppressed, and the Rb MOT begins to decay.  When the experiment is repeated with the push beam on, the
Rb MOT loading proceeds unimpeded, and the Cs MOT reaches $\sim70\%$ of its single-species level before the Rb and Cs clouds begin to overlap,
and interspecies inelastic collisions prevent further increase in the Cs atom number.  If necessary, the overlap could be reduced further by
displacing the Cs MOT as well.  Using the displaced MOT technique, we have created two-species Rb-Cs MOTs containing a total of more than $10^9$
atoms.

\subsection{Magnetic trap}
At the end of the MOT loading phase, a standard sequence is used to load the atoms into the magnetic trap.  A 30\,ms compressed MOT (CMOT) phase
and a 15\,ms optical molasses phase are employed to compress and further cool the atomic clouds. The position of the trapped clouds is adjusted
to be located at the centre of the magnetic trap by applying an appropriate bias field during the CMOT phase. Following the optical molasses,
the atoms are optically pumped into the $|F\!=\!3,m_F\!=\!-3\rangle$ and $|F=\!1,m_F\!=-1\rangle$ states of Cs and Rb respectively, with
efficiencies of $\geq65\%$ for both species. The push beam is switched off immediately before the start of the CMOT phase, so that the two
trapped atomic clouds become well overlapped prior to loading the magnetic trap. The short duration of the magnetic trap loading phase, however,
means that there are no detectable losses due to interspecies light assisted collisions even though the clouds are now well overlapped. We load
the atoms into the magnetic trap by simultaneously switching on a current of 150\,A in the baseball coil and 100\,A in the bias coil. This
produces a relatively weak trapping potential (see Table\,1) with a bias field of 166.0(3)\,G in order to reduce the potential energy imparted
to the cloud and to avoid the Cs Feshbach resonances that occur below $\sim135$\,G. When only a single species is collected in the science MOT,
typically up to 6(1)$\times10^8$ Rb or 3(1)$\times10^8$ Cs atoms can then be loaded into the magnetic trap.

The effect of the displaced MOT technique on the magnetic trap loading is vividly illustrated by absorption images of Rb and Cs atoms in the
magnetic trap.  Figures \ref{pushbeam}(b)-(e) contain images of magnetically trapped clouds taken with and without the push beam on during the
MOT phase.  The figure also contains vertical cross-sections of the atom clouds, showing the increase in optical depth for magnetically-trapped
atoms loaded from the displaced MOT. The flexibility of the displaced MOT technique allows us to vary the composition of the trapped mixture
over a wide range by adjusting the load duration of each species. We can load mixtures containing an equal amount of each species at close to
their respective maximum single-species atom numbers. At the other extreme, we can load one species to its single-species maximum together with
an arbitrarily small amount of the second species.

Preliminary observations show that the lifetime of the trapped atomic mixture in the magnetic trap at the load field of 166\,G is consistent
with the corresponding single-species lifetimes. This provides an excellent starting point for further experiments. Ongoing and future
experiments include the search for interspecies Feshbach resonances and investigations of sympathetic cooling of Cs with Rb in a tight, low bias
field trap.

\section{Conclusions}

In summary, we have demonstrated magnetic trapping of a mixture of cold Cs and Rb atoms in the $|F=3,m_F=-3\rangle$ and $|F=1, m_F=-1 \rangle$
states.  To our knowledge, this is the first time these states of Rb and Cs have been trapped together.  After measurements on the Rb-Cs science
MOT showed strong losses due to light-assisted inelastic collisions between the two species, a simple but effective technique was developed
which suppresses such collisions by spatially separating the Rb and Cs MOTs during loading.  This method is not specific to the Rb-Cs mixture,
and should prove useful in reducing loss in other two-species MOTs.  A particular advantage is that the push beam technique allows considerable
flexibility in the ratio of Rb to Cs atoms available for loading into the magnetic trap.  Our system thus represents an advantageous starting
point for a range of experiments on cold Rb-Cs mixtures, including a search for interspecies Feshbach resonances and studies of sympathetic
cooling.

\ack This work was funded by the Engineering and Physical Sciences Research Council (grant GR/S78339/01). SLC acknowledges the support of the
Royal Society. MLH acknowledges the Universities UK Overseas Research Scheme.  We thank C. J. Foot and D. Cassettari for their contributions to
the design of the pyramid MOT chamber, and I. G. Hughes for many helpful discussions.

\appendix
\section*{Appendix}
\setcounter{section}{1}

We derive the expression for $\beta_{ij}$ by starting with the rate equation for the two-species MOT
\begin{equation}
\frac{dN_{i}}{dt}=L-\gamma N_{i}-\beta_i \int_V n^2_{i}d^3r -\beta_{ij} \int_Vn_{i}n_{j}d^3r
\end{equation}
and recalling that the average densities for single and two-species atom clouds are defined as
\begin{eqnarray}
<\!\!n_i\!\!>=\frac{1}{N_i}\int_V n_i^2 d^3r \nonumber \\
<\!\!n_{ij}\!\!>=\frac{1}{N_i}\int_V n_in_j d^3r.
\end{eqnarray}
Using these definitions, the rate equation becomes
\begin{equation}
\frac{dN_{i}}{dt}=L-\gamma N_{i}-\beta_i <\!\!n_i\!\!>N_i - \beta_{ij}<\!\!n_{ij}\!\!>N_i.
\end{equation}
These definitions also allow us to formally define the `overlap' factor $F_{ij}$ as
\begin{equation}
F_{ij}=\frac{\int_V n_jn_i/N_id^3r}{\int_V n_in_i/N_id^3r}=\frac{<\!\!n_{ij}\!\!>}{<\!\!n_i\!\!>}.
\end{equation}

To obtain an expression for $\beta_{ij}$, steady-state solutions to the rate equation are found in the presence and absence of a second species.
In the single-species MOT, this gives
\begin{equation}
L=\gamma \tilde{N}_i + \beta_i <\!\!\tilde{n}_i\!\!> \tilde{N}_i
\end{equation}
where, as in the text, the $\tilde{\:}$ denotes quantities in the single-species MOT. Adding a second species changes the equation to
\begin{equation}
L=\gamma N_i + \beta_i <\!\!n_i\!\!>N_i + \beta_{ij} <\!\!n_{ij}\!\!>N_i.
\end{equation}
Solving for $\beta_{ij}$ gives the result:
\begin{equation}
\beta_{ij}= \frac{1}{<\!\!n_{ij}\!\!>N_i}\left[\gamma
\left(\tilde{N}_i-N_i\right)+\beta_i\left(<\!\!\tilde{n}_{i}\!\!>\tilde{N}_i-<\!\!n_{i}\!\!>N_i\right)\right].\label{eq:betaprimedensities}
\end{equation}

\section*{References}



\end{document}